\documentclass[fleqn,usenatbib]{mnras}

\usepackage{newtxtext,newtxmath}
\usepackage{CJK}
\usepackage{fontawesome}
\usepackage[colorinlistoftodos, textsize=tiny]{todonotes}

\usepackage{bm}
\usepackage{subcaption}
\usepackage{diagbox}
\usepackage{times}

\usepackage[english]{babel}
\usepackage[utf8x]{inputenc}

\usepackage{booktabs} 

\usepackage{amsmath,amsfonts}
\usepackage{graphicx}

\usepackage[T1]{fontenc}

\DeclareRobustCommand{\VAN}[3]{#2}
\newcommand{\vect}[1]{\mathbf{#1}}
\let\VANthebibliography\thebibliography
\def\thebibliography{\DeclareRobustCommand{\VAN}[3]{##3}\VANthebibliography}

\title[Cyclic-Permutation Invariant Networks for Classifying Periodic Variables]{Classification of Periodic Variable Stars with Novel Cyclic-Permutation Invariant Neural Networks}

\author[K. Zhang \& J. S. Bloom]{
Keming Zhang \begin{CJK*}{UTF8}{gkai}(张可名)\end{CJK*},$^{1}$\thanks{E-mail: kemingz@berkeley.edu}
Joshua S. Bloom$^{1,2}$
\\
$^{1}$Department of Astronomy, University of California, Berkeley, CA 94720-3411, USA\\
$^{2}$Lawrence Berkeley National Laboratory, 1 Cyclotron Road, MS 50B-4206, Berkeley, CA 94720, USA\\
}

\date{Accepted XXX. Received YYY; in original form ZZZ}

\pubyear{2020}

\begin{document}
\label{firstpage}
\pagerange{\pageref{firstpage}--\pageref{lastpage}}
\maketitle



\begin{abstract}
Neural networks (NNs) have been shown to be competitive against state-of-the-art feature engineering and random forest (RF) classification of periodic variable stars. Although previous work utilising NNs commonly operated on period-folded light-curves, no approach to date has taken advantage of the fact that network predictions should be invariant to the initial phase of the period-folded sequence. Initial phase is exogenous to the physical origin of the variability and should thus be immaterial to the downstream application. Here, we present cyclic-permutation invariant networks, a novel class of NNs for which the output is invariant to phase shifts by construction. We implement this invariance by means of ``Symmetry Padding.'' Across three different datasets of variable star light curves, we show that two implementations of the cyclic-permutation invariant network: the iTCN and the iResNet, consistently outperform non-invariant baselines and reduce overall error rates by between 4\% to 22\%. Over a 10-class OGLE-III sample, the iTCN/iResNet achieves an average per-class accuracy of 93.4\%/93.3\%, compared to RNN/RF accuracies of 70.5\%/89.5\% in a recent study using the same data.
Finding improvement on a non-astronomy benchmark, we suggest that the methodology introduced here should also be applicable to a wide range of science domains where periodic data abounds due to physical symmetries.
\end{abstract}


\section{Introduction}
Periodic variability arises across the Hertzsprung-Russell diagram and manifest through stellar pulsation, rotation, and/or binarity. The identification of dozens of distinct phenomenological sub-classes (e.g., \citealt{gaia_collaboration_gaia_2019}) reflects the richness of the underlying physical processes giving rise to observable changes in brightness and colour. Periodic variables can also serve as precision probes of distance \citep{1997eds..proc..273P}, line-of-sight dust extinction \citep{2008AJ....135..631K}, and Galactic structure \citep{Skowron478}. As such, the systematic discovery and classification of periodic variables in large time-domain surveys, some with billions of stars monitored, remains paramount.

At scale, human expert labelling of variability catalogues of light curves has naturally, in the past decade, given way to automated classification approaches with machine learning. Random forest (RF; \citealt{random_forest}) classification, while performant, requires computationally expensive, hand-crafted feature engineering as part of data preprocessing \citep{richards_machine-learned_2011,kim_package_2016}. More recently, deep representation learning has further pushed the boundaries by learning not only decision rules on features of raw data, but also the low-dimensional feature representation itself. This approach has advanced many fields in astronomy (e.g., \citealt{kim_stargalaxy_2017,shallue_identifying_2018,agarwal_fetch_2020,zhang_deepcr_2020}).

For variable star classification, both convolutional neural networks (CNNs; \citealt{lecun_deep_2015}) and recurrent neural networks (RNNs; \citealt{hochreiter_long_1997,cho_learning_2014}) have been shown to be competitive to the traditional RF-based methods. \cite{naul_recurrent_2018} used an RNN autoencoder network to learn low-dimensional representations of period-folded light-curves in an unsupervised fashion. This representation was then, in a supervised context, used as feature inputs to a RF classifier. They showed that the learned features are at least as good as, and often better than, two sets of state-of-the-art hand-crafted features \citep{richards_machine-learned_2011,kim_package_2016}, in terms of downstream classification accuracy. \cite{becker_scalable_2020} used an RNN for which instead of period-folding, each input light curve is grouped with a moving window of size 50 and stride 25. Although period-folding improves performance \citep{naul_recurrent_2018}, \cite{becker_scalable_2020}'s time-space RNN does not require the period to be calculated, and is thus less computationally expensive in terms of preprocessing. Again, they found similar performance to a RF classifier with the \cite{nun_fats_2015} features over three datasets, although lower accuracy was seen for many sub-classes with the OGLE dataset (Table \ref{tab:classification_ogle}; see Section \ref{sec:ogle} for data description). More recently, \cite{jamal_neural_2020} systematically benchmarked the performance of different configurations of RNN and CNN network architectures on variable star classification. Aside from other work (e.g., \citealt{tsang_deep_2019,aguirre_deep_2018}) evaluating neural network performance retrospectively on previously labeled datasets, \cite{dekany_near-infrared_2020} used an RNN classifier to identify a new sample of fundamental-mode RR Lyrae (RRab) stars. Similarly, \cite{dekany_into_2019} found Classical and Type II Cepheids with a CNN classifier, also using the VISTA Variables in the Via Lactea (VVV) survey \citep{minniti_vista_2010} and using period-folded light curves.

While the compact phase-space (i.e., period-folded) light-curve representation has been widely adapted in the aforementioned studies, none of the neural networks used therein guarantees the same prediction under phase-shifts, or cyclic-permutations, of the same period-folded light curve. Since the initial-phase of the phase-space sequence is experimentally determined and exogenous to the physical origin of the variability, it could be seen as a nuisance parameter which should not affect classification. In the limit where a classification task is non-trivial---either due to the inherent difficulty of class separability or low signal-to-noise data---some degree of domain knowledge can, in principle, be injected into the network architecture through known symmetries and conservation laws \citep{2019RvMP...91d5002C,mattheakis_physical_2020}. Neural networks for computer vision tasks, for example, have been developed that are scale, rotation, and translation invariant \citep{2015arXiv150602025J}. Specialised networks for particle physics inference preserve known properties of quantum chromodynamics \citep{2019JHEP...01..057L}. For periodic time series, we seek a network architecture with built-in invariance to cyclic-permutation to improve performance.

Here, we present cyclic-permutation invariant convolutional networks. We describe specific implementations with 1-D residual convolutional networks (ResNets), and with dilated 1-D convolutional networks (TCN) that have been shown to achieve state-of-the-art for a variety of sequence modeling tasks \citep{bai_empirical_2018}. The cyclic-permutation invariant network is descried in Section \ref{sec:method}, whereas the variable star datasets used in benchmarking the invariant network against previous methods are discussed in Section \ref{sec:data}. Finally, the performance of the invariant networks in various scenarios are discussed in Section \ref{sec:results}. To facilitate applications of the cyclic-permutation invariant networks, we are releasing code at \url{https://github.com/kmzzhang/periodicnetwork}.

\begin{figure*}
\begin{center}
 \includegraphics[width=\textwidth]{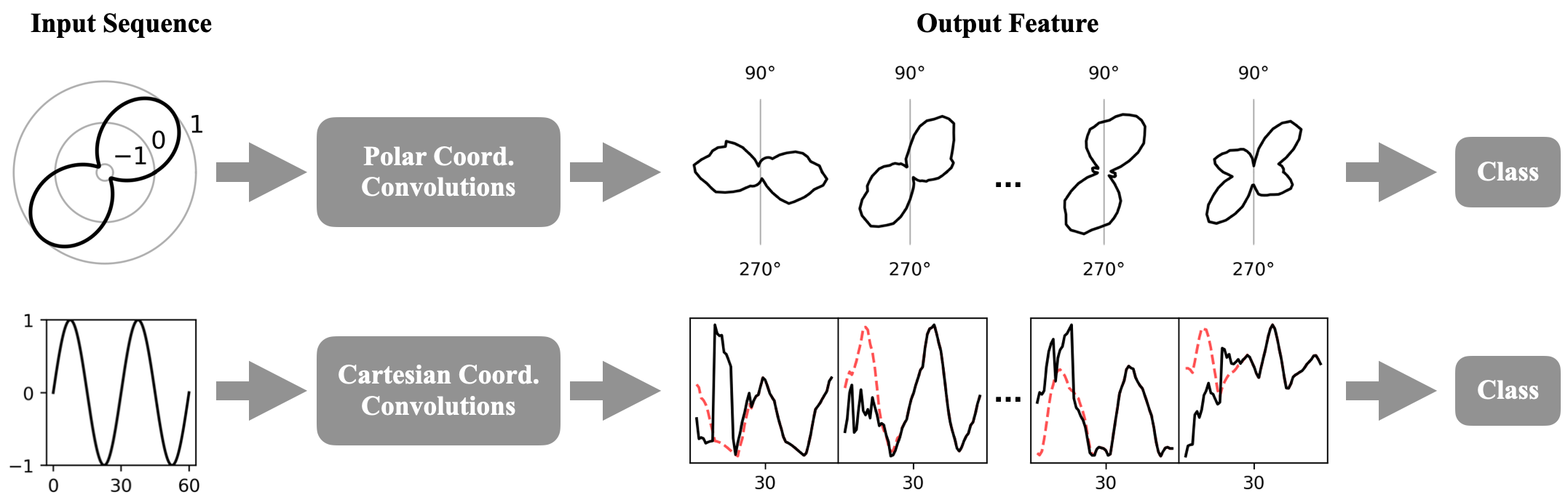}
 \end{center}
 \caption{Schematic illustration of the effect of polar coordinate convolutions in preserving cyclic-permutation invariance. The input and output sequences are shown in polar coordinates for iTCN (top), and in Cartesian coordinates for TCN (bottom). The input sequence is a sine curve with two full oscillations in both cases. In the upper diagram, 1-D feature maps of the periodic input remains periodic; rotational symmetry is preserved. These periodic feature maps are also shown in Cartesian coordinates of the lower plots in red dashed lines for comparison. As demonstrated by the discrepancy, feature maps are distorted for the first full oscillation in the non-invariant network, which is shown in solid black lines.
 }
 \label{fig:phase}
\end{figure*}

\begin{figure*}
 \centering
 \includegraphics[width=\textwidth]{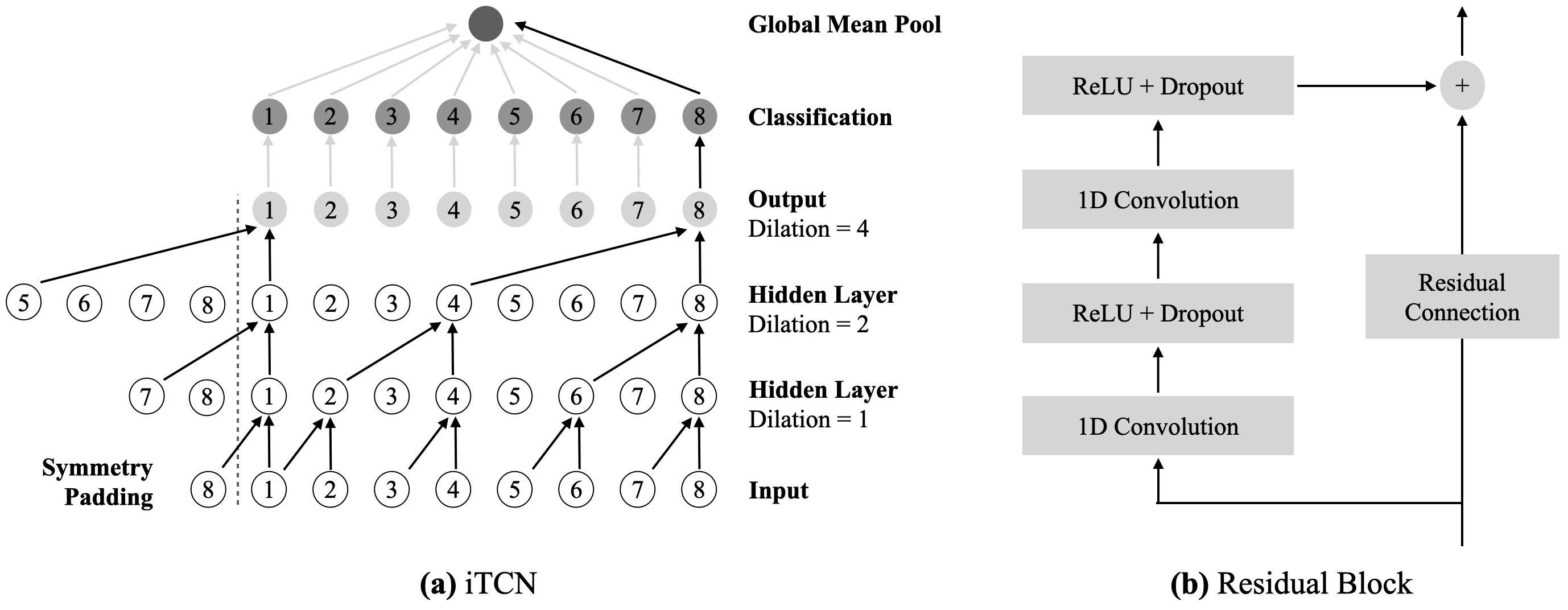}
 \caption{(a) Simplified illustration of the cyclic-permutation invariant Temporal Convolutional Network (iTCN). Numbers refer to the ordering of the period-folded sequence. 
 Dilated convolutions are represented by arrows where the dilation factor is indicated to the right of each layer. Gray arrows in the final two layers represent operations which are present only in the iTCN not the TCN. The classification layer consists of two convolutions of \texttt{kernel} size 1. (b) The residual block, which is the actual hidden layer used in the iTCN. Residual connections are to be replaced with $k=1$ convolutions when consecutive layers have different hidden dimensions.}
 \label{fig:network}
\end{figure*}

\section{Method}
\label{sec:method}

The cyclic-permutation invariant networks that we introduce here refer to any neural network satisfying the following condition. Given an input sequence $\vect{x} \in \rm \mathbb{R}^{N}$, a neural network $f: \vect{x}\to \vect{y}$ is invariant to cyclic-permutations if
\begin{equation}
    \forall i\in [2, N], f(\vect{x}_{1:N}) = f({\rm concat}(\vect{x}_{i:N}, \vect{x}_{1:i-1}))
\end{equation}

We first offer a high-level overview of cyclic-permutation invariant networks before discussing implementation details. Under the cyclic-permutation invariant network framework, the multi-cycle periodic time series is first period-folded into a single cycle by transforming from temporal space ($\vect{t}, \vect{m}$) into phase space ($\pmb{\phi}, \vect{m}$): $\pmb{\phi}=\vect{t} \mod p$, where $m_{i}$ is the magnitude (or flux) measurement at phase $\phi_{i}$ and $p$, the period, is determined with periodogram analysis \citep{lomb_least-squares_1976,scargle_studies_1982}. The period is first used to fold the light-curve into phase-space and then concatenated to the output of the last convolution layer as an auxiliary input. We then make the observation that under polar coordinates, the period-folded sequence is essentially wrapped in a ``closed ring'' (Figure \ref{fig:phase}: Input Sequence, top row) where phase shifts simply become ``rotations'' which allows outputs to remain periodic (Figure \ref{fig:phase}: output feature). Phase-averaging the output feature map then results in a feature vector that is invariant to the initial phase ($\phi_0$), rendering it a nuisance parameter. On the other hand, for the usual Cartesian-coordinate CNNs, phase shifts result in different input sequences and therefore different outputs. Polar coordinate convolution is implemented by replacing zero-padding of length (\texttt{kernel} size $-$ 1) in ordinary Cartesian-coordinate CNNs with ``Symmetry Padding,'' which pads the input or hidden sequence not with zeros, but with the sequence itself (Figure \ref{fig:network}a).

Based on the above framework, we present two particular implementations of the cyclic-permutation invariant network: the invariant Temporal Convolutional Network (iTCN) and the invariant Residual Convolutional Network (iResNet). The iTCN is based on the Temporal Convolutional Network (TCN; \cite{bai_empirical_2018}), and is composed of ``residual blocks'' (Figure \ref{fig:network}b) of 1D dilated convolutions (Figure \ref{fig:network}a), where the input to each ``residual block' is concatenated to the output, creating a ``gradient highway'' for back-propagation, thus allowing for improved network optimization. Dilated convolutions refer to convolutions where the convolution \texttt{kernel} is applied over a region larger than the \texttt{kernel} size by skipping input values with a step of $2^{n-1}$ for the \hbox{$n$-th} layer. This dilation allows the network to achieve an exponential increase in the receptive field --- the extent of input data accessible with respect to a particular output neuron --- with network depth. The receptive field is calculated as
\begin{equation}
{\mathcal R} =  (K-1)\times \sum_{n=1}^{D} 2\times 2^{n-1} = (K-1) \times(2^{D+1} - 2), \label{eq:receptive}
\end{equation}
where $K$ is the kernel size, $D$ the number of layers, $2^{n-1}$ the dilation factor for the $n^{th}$ layer, and the additional factor of 2 due to the fact that each residual block consists of two dilated convolutions. Network depth is required to be large enough for the receptive field to be larger than the input sequence length, such that each feature vector in the output layer has complete information over the input sequence.
Simultaneous predictions are then made for every possible initial phase of the input sequence (Figure \ref{fig:network}a: ``classification'' layer) by first concatenating the period to each vector in the output layer, which serves as the feature vector for each phase. Each feature vector is then fed into a simple 2-layer feed-forward network that returns a vector with the same dimension as the number of classes. The outputs for the different phases are finally averaged with a \texttt{global mean pooling} layer as input to the \texttt{softmax} function for normalized class probabilities. By averaging predictions from all possible initial phases, the invariant network makes more robust predictions, as compared to non-invariant CNNs and RNNs, which can only predict for one particular initial phase with one network forward pass.

As a demonstration, for the toy iTCN network shown in Figure \ref{fig:network}a, the last time-step of the output sequence (gray circle ``8''; forth row bottom to top) is connected by arrows across the layers to the first time-step of the input sequence, and therefore has a receptive field of ${\mathcal R}=8$. Applying a cyclic-permutation to the input sequence (e.g. 2, 3, 4, 5, 6, 7, 8, 1) would result in the same cyclic-permutation to the output sequence, which does not change the final classification because classification from each time-step is averaged, thus making the network invariant to such permutations.

To visualise the effects of cyclic-permutation invariance on modelling periodic sequences, we compare output sequences produced by the iTCN and the TCN in Figure \ref{fig:phase}. We create an iTCN and a TCN with the same weights and the same receptive field of ${\mathcal R}=30$ at a network depth of 4 with kernel size 2. The input sequence is a length-60 sine function with two full oscillations (0 to $4\pi$ radian). As seen in the figure, while output feature maps produced by the iTCN remain symmetrical in polar coordinates, the first half of the output sequence produced by the TCN is distorted by zero-padding, thus degrading the fidelity of output feature maps.

The second implementation, the iResNet, is also composed of stacks of ``residual blocks,'' but is different from the iTCN in that the exponential receptive-field increase is achieved through \texttt{max pooling} layers, instead of dilated convolutions. A \texttt{max pooling} layer of \texttt{kernel} size 2 and \texttt{stride} 2, which combines every two adjacent feature vectors into one by selecting the maximum value, is added after every ``residual block'' to distil information extracted. After every such operation, the temporal dimension is reduced by half, while the number of hidden dimension is doubled until a specified upper limit. Unlike the iTCN, the feature vectors in the iResNet output layer do not have a one-to-one correspondence to the input sequence because the temporal dimension of the output feature map is reduced by a factor of $2^{D-1}$, where D is the network depth. Because of this discreteness of featurisation, the iResNet is only invariant to phase shifts of $2^{D-1}$ steps (when the input sequence length is divisible by the same factor). Nevertheless, the iResNet potentially benefit from data augmentation of the input-sequence initial phase, as could non-invariant networks (see Appendix \ref{sec:aug}).

\section{Benchmark Data}
We assembled benchmarking datasets from three publicly available datasets of variable star light curves: All-Sky Automated Survey for Supernovae (ASAS-SN; \citealt{jayasinghe_asas-sn_2018,jayasinghe_asas-sn_2019}), Massive Compact Halo Object (MACHO; \citealt{alcock_macho_1996}), and Optical Gravitational Lensing Experiment (OGLE-III; \citealt{udalski_optical_2003}). The datasets are described below.
\label{sec:data}
\subsection{All-Sky Automated Survey for Supernovae (ASAS-SN) Data}
The ASAS-SN dataset consists of 282,795 light curves from eight classes of variable stars: 288 W\,Virginis ($p >$ 8 day), 102 W\,Virginis ($p < 8$ day), 941 Classical Cepheids, 297 Classical Cepheids (Symmetrical), 1,631 Delta Scuti, 25,314
Detached Eclipsing Binaries, 12,601
Beta Lyrae, 43,151
W Ursae Majoris-type, 2,149
High Amplitude Delta Scuti, 9,623
Delta Scuti, 14046
Rotational Variables, 26,956
RR Lyrae type A/B, 7,469
RR Lyrae type C, 364
RR Lyrae type D, and 137,847
Semi-regular Variables.
The class label of each variable star is classified by \citet{jayasinghe_asas-sn_2019} and only those with class probability greater than $99\%$ are used. The maximum number of full light curve per class is capped at 20,000 to reduce the number of light curves of the dominant classes. Finally, segmenting into $L=200$ chunks results in 106,005 fixed-length light curves.

\subsection{Massive Compact Halo Object (MACHO) Project data}
The MACHO dataset, taken directly from \cite{naul_recurrent_2018}, consists of 21,470 red band light curves from eight classes of variable stars: 7,403 RR Lyrae AB, 6,833 Eclipsing Binary, 3,049 Long-Period Variable Wood (sub-classes A--D were combined into a single super-class), 1,765 RR Lyrae C, 1,185 Cepheid Fundamental, 683 Cepheid First Overtone, 315 RR Lyrae E, and 237 RR Lyrae/GB Blend. Segmenting into $L=200$ chunks has resulted in 80,668 fixed-length light curves.

\subsection{Optical Gravitational Lensing Experiment: OGLE-III}
\label{sec:ogle}
The OGLE-III dataset is identical to that used in \cite{becker_scalable_2020}, except for the selection of OSARGs. The OGLE-III data consists of 357,748 light curves from ten classes of variable stars:
6862 Eclipsing Contact Binaries,
21503 Eclipsing Detached Binaries, 
9475 Eclipsing Semi-detached Binaries, 
6090 Miras,
234,932 OGLE Small Amplitude Red Giants (OSARG),
25943 RR Lyrae type A/B,
7990 RR Lyrae type C,
34835 Semi-regular Variables,
7836 Classical Cepheids,
2822 Delta Scuti. Of the 234,932 OSARGs, 40,000 random ones are selected. Absent of a fixed random seed in their relevant code section, we have not been able to procure their exact selection, although the number selected is large enough for the difference to be small. Finally, segmenting into $L=200$ chunks results in 540,457 fixed-length light-curves.

\section{Results}
\label{sec:results}

\begin{table}
    \begin{tabular}[\columnwidth]{*{4}{c}}
    \toprule
    Model & MACHO& OGLE-III& ASAS-SN\\
    \midrule
    iTCN& $\textbf{92.7\%}\pm0.43\%$& $\textbf{93.7\%}\pm0.09\%$& $\textbf{94.5\%}\pm0.14\%$ \\
    TCN& $92.0\%\pm0.35\%$& $92.9\%\pm0.11\%$& $93.0\%\pm0.20\%$ \\
    $\textit{diff }^{1}$ & $({-0.58\%}_{-0.17\%}^{+0.05\%})$& $({-0.75\%}_{-0.02\%}^{+0.04\%})$& $({-1.52\%}_{-0.09\%}^{+0.12\%})$ \\
    \midrule
    iResNet$^{*}$ & $92.6\%\pm0.45\%$& $\textbf{93.7\%}\pm0.09\%$& $93.9\%\pm0.14\%$ \\
    ResNet& $92.1\%\pm0.34\%$& $93.4\%\pm0.11\%$& $93.2\%\pm0.22\%$ \\
    $\textit{diff }^1$ & $({-0.48\%}_{-0.13\%}^{+0.05\%})$& $({-0.25\%}_{-0.02\%}^{+0.03\%})$& $({-0.64\%}_{-0.07\%}^{+0.01\%})$ \\
    \midrule
    \midrule
    GRU& $92.3\%\pm0.37\%$& $92.8\%\pm0.19\%$& $93.6\%\pm0.42\%$ \\
    $\textit{diff }^2$ & $({-0.34\%}_{-0.02\%}^{+0.05\%})$& $({-0.86\%}_{-0.13\%}^{+0.17\%})$& $({-0.71\%}_{-0.12\%}^{+0.07\%})$ \\
    LSTM& $91.7\%\pm0.53\%$& $92.6\%\pm0.61\%$& $93.5\%\pm0.23\%$ \\
    $\textit{diff }^2$ & $({-0.93\%}_{-0.16\%}^{+0.45\%})$& $({-0.85\%}_{-0.48\%}^{+0.17\%})$& $({-1.00\%}_{-0.06\%}^{+0.11\%})$ \\
    \bottomrule
 \end{tabular}
  \caption{\label{table:vs} {\bf Ablation study test accuracies demonstrating gains afforded by cyclic-permutation invariance}. The network with the top accuracy for each dataset is shown in bold. Test accuracies are the mean values for 8 different data splits. Median test accuracy differences of the different data partitions are shown in parentheses with the uncertainty interval corresponding to 1-$\sigma$ range of test accuracy differences calculated pair-wise for the same random partitions of data. Negative accuracy differences indicate better performances of the invariant network.\\
  $^{1}$Compared to the invariant version of the same network.\\
  $^{2}$Compared to the best performing network.\\
  $^{*}$Semi-invariant due to use of discrete max-pooling layers.}
\end{table}

We first study the evaluation metrics of the iTCN/iResNet architectures compared to the TCN/ResNet architectures to identify the gains made solely by cyclic-permutation invariance. Such ``ablation'' studies---applying a single change to the network architecture during training and testing to isolate the effect of that change---are common in deep learning. Input data is given in phase-space in all cases as the advantages compared to time-space have been demonstrated in previous studies (e.g. \citealt{naul_recurrent_2018}). RNN baselines of GRU \citep{cho_learning_2014} and LSTM \citep{hochreiter_long_1997} are also included as additional baseline methods for comparison. We also note that cyclic-permutation invariance is forbidden in RNNs because of its acyclic topology.

The networks are trained on fixed-length light-curve segments ($L=200$) from the three datasets described in Section \ref{sec:data}. We apply randomised, stratified 60/20/20 train/validation/test splits for each dataset. To properly account for dataset boot-strapping noise---accuracy variations due to the particular choices of data splits---the same random splits are used to test every network, whose accuracies are compared pairwise in splits. We emphasise that the standard deviation of the accuracy differences, rather than the standard deviation of the accuracy themselves, should then serve as the basis for comparison of the accuracies. This is because accuracy variations for a given network and dataset are largely dominated by the boot-strapping noise due to train/test partitioning, and thus would be an overestimation of the variances solely attributed to the networks. Within each split, each full light-curve is divided into sequences of length 200 in temporal order and transformed into phase-space with respect to the period provided in the catalogs. Compared to random sampling, subdividing by temporal order preserves the irregular samplings which resemble how data is accumulated. Each segment is then individually normalised (zero-mean and unit variance) while measurement times are rescaled by the period into phase ([0, 1]). The measurement phase intervals $\Delta\pmb{\phi}$ between successive data points are fed together with the rescaled light curve as inputs to the network. The mean and standard deviation of each light curve segment, along with $\log(p)$, are concatenated to the network output layer as auxiliary inputs. We perform extensive hyperparameter optimisation for each pair of network and dataset (see Appendix \ref{sec:hyperops}).

As shown in Table \ref{table:vs}, the improvements of the iTCN and the iResNet from their respective non-invariant baselines, as well as the RNNs, are significant by more than 5-$\sigma$ in most cases, demonstrating the advantages of enforcing cyclic-permutation invariance. The improvements in classification accuracies correspond to reductions in overall error rates by between 4\% to 22\%, depending upon the non-invariant baseline and the dataset.

\subsection{Comparison to published methods and results}
\label{sec:comparison}
We first consider the time-space RNN and RF results recently published in \cite{becker_scalable_2020}. \cite{becker_scalable_2020} presents OGLE-III classification results with their time-space GRU and an RF baseline with the \cite{nun_fats_2015} features. The Becker et al.~time-space GRU work groups each full OGLE-III light curve with a moving window of size 50 and stride 25, whereby the effective sequence length is reduced by a factor of 25. This reduction alleviates the so-called vanishing gradient problem \citep{hochreiter_long_1997} which limits the sequence length that the RNN could be effectively trained on. To facilitate this comparison, we have used the same OGLE-III data selection as their work (Section \ref{sec:ogle}). Since a $L>300$ requirement has been applied to their OGLE-III data selection, we trained the iTCN/iResNet on $L=300$ segments, and average classifications on $L=300$ segments for each full light curve during testing. As seen in Table \ref{tab:classification_ogle}, the cyclic-permutation invariant networks outperform both results. The invariant network accuracies are significantly higher for most classes, reducing error rates by as much as 69\% for the minority classes. We find this result to be critically important, as the hard-to-classify minority classes tend to be the least well-understood and often are the most interesting to identify for further study. In particular, the largest error rate reductions against RF are seen in Eclipsing Binaries, Delta Scuti, and Semi-Regular Variables, which are important both for accurate tests of stellar evolution models (e.g. \citealt{guinan_eclipsing_2000,torres_absolute_2002}) and for precision probes of distance (e.g. \citealt{bonanos_first_2006,mcnamara_2007,north_vlt_2012}).

Additionally, \cite{naul_recurrent_2018} published RF benchmark accuracies for the MACHO dataset of 90.50\% with the \cite{richards_machine-learned_2011} features and 88.98\% with the \cite{kim_package_2016} features. While we have use the same MACHO dataset as \cite{naul_recurrent_2018}, our results are not directly comparable because \cite{naul_recurrent_2018} preformed randomised train/test split on the $L=200$ segmented light curves, which have caused different versions of the same light curve to exist in both training and test split, resulting in information leakage and thus a higher accuracy.

\begin{table}
    \centering
    \begin{tabular}[width=\linewidth]{l|cc|cc} 
        \hline
        Class &iTCN &iResNet & GRU & RF \\
        \hline
        Cep & \textbf{98.3\%}$\pm$0.3\% & \textbf{98.4\%}$\pm$0.7\% & 72\% & 97\%\\
        RRab & \textbf{99.7\%}$\pm$0.1\% & \textbf{99.7\%}$\pm$0.4\%& 85\% & 99\%\\
        RRc & \textbf{99.0\%}$\pm$0.2\% & \textbf{99.1\%}$\pm$0.1\%& 30\%&  98\%\\
        Dsct& \textbf{97.6\%}$\pm$0.8\% & \textbf{97.8\%}$\pm$0.6\%& 72\%& 93\%\\
        EC  & \textbf{87.9\%}$\pm$0.9\% & \textbf{87.8\%}$\pm$0.7\%& 54\% & 79\%\\
        ED  & \textbf{95.0\%}$\pm$0.3\% & \textbf{94.8\%}$\pm$0.4\%& 93\% & 92\%\\
        ESD & 68.7\%$\pm$1.0\% & \textbf{70.7\%}$\pm$0.9\%& 24\% & 61\%\\
        Mira & \textbf{97.1\%}$\pm$0.6\% &\textbf{96.8\%}$\pm$0.3\% & 92\% & \textbf{97\%}\\
        SRV & \textbf{96.0\%}$\pm$0.4\% & \textbf{95.9\%}$\pm$0.2\%& 93\% & 82\%\\
        OSARG & 93.2\%$\pm$0.4\% &93.4\%$\pm$0.2\% & 90\%& \textbf{97\%}\\
        \hline
        Mean & \textbf{93.4\%} &\textbf{93.3\%} & 70.5\%& 89.5\%\\
        \hline
    \end{tabular}
    \caption{\label{tab:classification_ogle} {\bf Test accuracies for OGLE-III full-length light curves compared to classifications results in} \protect\cite{becker_scalable_2020}. For all but one subclass, the cyclic-permutation invariant networks outperform previous results. Similar to Table \ref{table:vs}, we note that uncertainties are dominated by the bootstrapping noise arising from randomized data partitioning, and as such, are only upper limits to uncertainties in the accuracy differences for each class.}
\end{table}

\subsection{Adapting to variable-length sequences}
\label{sec:adaptive}

Although none of the networks tested are restricted to fixed-length inputs, we emphasise that fixed-length sequence trained networks should not be naively applied to test sequences of different lengths because doing so results in degraded accuracy: different sequence lengths correspond to a different effective sampling frequency in phase space. The neural network is essentially asked to extrapolate, not interpolate, beyond the training function domain.

\begin{figure}
 \centering
 \includegraphics[width=\columnwidth]{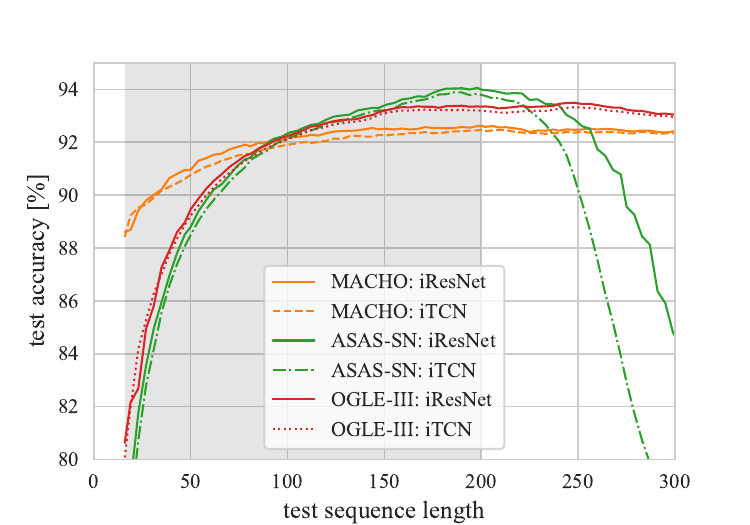}
 \caption{iResNet/iTCN test accuracy as a function of test sequence length for MACHO, ASAS-SN, and OGLE-III. Shaded region indicates the range of training sequence length: $16<L<200$}
 \label{fig:varlen}
\end{figure}

In Table \ref{tab:classification_ogle}, we showed a segment-and-classify scheme which is shown to be effective for OGLE-III full light curves. Here, we provide examples to show how the invariant networks could be directly trained on variable length sequences. A random sequence length in between $16<L<200$ is selected for each mini-batch during training. The optimal hyper-parameters for $L=200$ networks are used, though each network could potentially benefit from increased complexity due to the increased task difficulty. As seen in Figure \ref{fig:varlen}, high accuracy is maintained across a wide range of sequence lengths within the training range of $16<L<200$. Beyond the training range $16<L<200$, the ability of the networks to generalise is dataset dependent.

Furthermore, we note that the optimal range of training sequence length depends on the ratio of the period to the cadence. If the cadence is short compared to the periods, then the training sequence length should have a longer upper limit for each training light curve to cover at least one oscillation period. Figure \ref{fig:varlen} also suggests a way by which the training sequence length upper limit could be determined. As accuracy only increase marginally for MACHO beyond $L\sim100$, a shorter upper limit could be selected whereby each full-length light curves is cut into more segments whose results are combined. On the other hand, the training sequence length upper limit could be increased for ASAS-SN, as classification accuracy is still on the rise at $L=200$, which suggests that the networks are still gaining additional information with increasing sequence length near the $L=200$ cutoff.

\section{Conclusions}
Large scale time-domain surveys have both generated the need for, and enabled the training of, effective data-driven classification techniques for both periodic and non-periodic variable sources.
In this work, as in other fields with established benchmark datasets, we have decoupled methodology from data and shown that the cyclic-permutation invariant networks achieve state-of-the-art accuracies for periodic variable star classification on datasets previously acquired. While the networks perform well on light curves with few data-points, we did not test the efficacy of such networks in a streaming context, where the period is not known \textit{a priori.}  Future work could explore how the invariant networks can be used in a streaming context, as well as efficient neural and non-neural ML methods for non-periodic data \citep{tachibana_deep_2020,moller_supernnova_2020,narayan_machine-learning-based_2018}, which when combined with the methodology for periodic sources introduced here, can serve as the basis of a generalised classification framework for modern time-domain surveys.

\section*{Data Availability}
We have made code for reproducing the above experiments publicly available at
\url{https://github.com/kmzzhang/periodicnetwork}. All three variable star light curve datasets have been made publicly available on Zenodo. Please refer to the GitHub repository for download instructions and usage.

\section*{Acknowledgements}
K.Z.\ and J.S.B.\ are supported by a Gordon and Betty Moore Foundation Data-Driven Discovery grant. J.S.B.\ is partially sponsored by a faculty research award from Two Sigma. K.Z.\ thanks the LSSTC Data Science Fellowship Program, which is funded by LSSTC, NSF Cybertraining Grant 1829740, the Brinson Foundation, and the Moore Foundation; his participation in the program has benefited this work. We thank Benny T.\ H.\ Tsang for assistance with ASAS-SN data and Jorge Mart\'inez-Palomera for assistance with OGLE-III data. We thank Sara Jamal for comments on a draft of the manuscript. This work is supported by the Amazon Web Services (AWS) Cloud Credits for Research program.



\begin{thebibliography}{}
\makeatletter
\relax
\def\mn@urlcharsother{\let\do\@makeother \do\$\do\&\do\#\do\^\do\_\do\%\do\~}
\def\mn@doi{\begingroup\mn@urlcharsother \@ifnextchar [ {\mn@doi@}
  {\mn@doi@[]}}
\def\mn@doi@[#1]#2{\def\@tempa{#1}\ifx\@tempa\@empty \href
  {http://dx.doi.org/#2} {doi:#2}\else \href {http://dx.doi.org/#2} {#1}\fi
  \endgroup}
\def\mn@eprint#1#2{\mn@eprint@#1:#2::\@nil}
\def\mn@eprint@arXiv#1{\href {http://arxiv.org/abs/#1} {{\tt arXiv:#1}}}
\def\mn@eprint@dblp#1{\href {http://dblp.uni-trier.de/rec/bibtex/#1.xml}
  {dblp:#1}}
\def\mn@eprint@#1:#2:#3:#4\@nil{\def\@tempa {#1}\def\@tempb {#2}\def\@tempc
  {#3}\ifx \@tempc \@empty \let \@tempc \@tempb \let \@tempb \@tempa \fi \ifx
  \@tempb \@empty \def\@tempb {arXiv}\fi \@ifundefined
  {mn@eprint@\@tempb}{\@tempb:\@tempc}{\expandafter \expandafter \csname
  mn@eprint@\@tempb\endcsname \expandafter{\@tempc}}}

\bibitem[\protect\citeauthoryear{Agarwal, Aggarwal, Burke-Spolaor, Lorimer  \&
  Garver-Daniels}{Agarwal et~al.}{2020}]{agarwal_fetch_2020}
Agarwal D.,  Aggarwal K.,  Burke-Spolaor S.,  Lorimer D.~R.,   Garver-Daniels
  N.,  2020, \mn@doi [Monthly Notices of the Royal Astronomical Society]
  {10.1093/mnras/staa1856}, 497, 1661

\bibitem[\protect\citeauthoryear{Aguirre, Pichara  \& Becker}{Aguirre
  et~al.}{2018}]{aguirre_deep_2018}
Aguirre C.,  Pichara K.,   Becker I.,  2018, \mn@doi [Monthly Notices of the
  Royal Astronomical Society] {10.1093/mnras/sty2836}, 482, 5078

\bibitem[\protect\citeauthoryear{Alcock et~al.,}{Alcock
  et~al.}{1996}]{alcock_macho_1996}
Alcock C.,  et~al., 1996, \mn@doi [The Astronomical Journal] {10.1086/117859},
  111, 1146

\bibitem[\protect\citeauthoryear{Arjovsky, Shah  \& Bengio}{Arjovsky
  et~al.}{2016}]{arjovsky_unitary_2016}
Arjovsky M.,  Shah A.,   Bengio Y.,  2016, arXiv:1511.06464 [cs, stat]

\bibitem[\protect\citeauthoryear{Bai, Kolter  \& Koltun}{Bai
  et~al.}{2018}]{bai_empirical_2018}
Bai S.,  Kolter J.~Z.,   Koltun V.,  2018, arXiv:1803.01271 [cs]

\bibitem[\protect\citeauthoryear{Becker, Pichara, Catelan, Protopapas, Aguirre
  \& Nikzat}{Becker et~al.}{2020}]{becker_scalable_2020}
Becker I.,  Pichara K.,  Catelan M.,  Protopapas P.,  Aguirre C.,   Nikzat F.,
  2020, \mn@doi [Monthly Notices of the Royal Astronomical Society]
  {10.1093/mnras/staa350}

\bibitem[\protect\citeauthoryear{Bonanos et~al.,}{Bonanos
  et~al.}{2006}]{bonanos_first_2006}
Bonanos A.~Z.,  et~al., 2006, \mn@doi [The Astrophysical Journal]
  {10.1086/508140}, 652, 313

\bibitem[\protect\citeauthoryear{Breiman}{Breiman}{2001}]{random_forest}
Breiman L.,  2001, Machine learning, pp 5--32

\bibitem[\protect\citeauthoryear{{Carleo}, {Cirac}, {Cranmer}, {Daudet},
  {Schuld}, {Tishby}, {Vogt-Maranto}  \& {Zdeborov{\'a}}}{{Carleo}
  et~al.}{2019}]{2019RvMP...91d5002C}
{Carleo} G.,  {Cirac} I.,  {Cranmer} K.,  {Daudet} L.,  {Schuld} M.,  {Tishby}
  N.,  {Vogt-Maranto} L.,   {Zdeborov{\'a}} L.,  2019, \mn@doi [Reviews of
  Modern Physics] {10.1103/RevModPhys.91.045002}, \href
  {https://journals.aps.org/rmp/abstract/10.1103/RevModPhys.91.045002} {91,
  045002}

\bibitem[\protect\citeauthoryear{Cho, van Merri\"enboer, Gulcehre, Bahdanau,
  Bougares, Schwenk  \& Bengio}{Cho et~al.}{2014}]{cho_learning_2014}
Cho K.,  van Merri\"enboer B.,  Gulcehre C.,  Bahdanau D.,  Bougares F.,
  Schwenk H.,   Bengio Y.,  2014, in Proceedings of the 2014 {Conference} on
  {Empirical} {Methods} in {Natural} {Language} {Processing} ({EMNLP}).
  Association for Computational Linguistics, Doha, Qatar, pp 1724--1734,
  \mn@doi{10.3115/v1/D14-1179}, \url
  {https://www.aclweb.org/anthology/D14-1179}

\bibitem[\protect\citeauthoryear{D\'ek\'any \& Grebel}{D\'ek\'any \&
  Grebel}{2020}]{dekany_near-infrared_2020}
D\'ek\'any I.,  Grebel E.~K.,  2020, \mn@doi [The Astrophysical Journal]
  {10.3847/1538-4357/ab9d87}, 898, 46

\bibitem[\protect\citeauthoryear{D\'ek\'any, Hajdu, Grebel  \&
  Catelan}{D\'ek\'any et~al.}{2019}]{dekany_into_2019}
D\'ek\'any I.,  Hajdu G.,  Grebel E.~K.,   Catelan M.,  2019, \mn@doi [The
  Astrophysical Journal] {10.3847/1538-4357/ab3b60}, 883, 58

\bibitem[\protect\citeauthoryear{{Gaia Collaboration} et~al.,}{{Gaia
  Collaboration} et~al.}{2019}]{gaia_collaboration_gaia_2019}
{Gaia Collaboration} et~al., 2019, \mn@doi [Astronomy and Astrophysics]
  {10.1051/0004-6361/201833304}, 623, A110

\bibitem[\protect\citeauthoryear{Guinan, Ribas, Fitzpatrick, Gim\'enez, Jordi,
  McCook  \& Popper}{Guinan et~al.}{2000}]{guinan_eclipsing_2000}
Guinan E.~F.,  Ribas I.,  Fitzpatrick E.~L.,  Gim\'enez \'A.,  Jordi C.,  McCook
  G.~P.,   Popper D.~M.,  2000, \mn@doi [The Astrophysical Journal]
  {10.1086/317211}, 544, 409

\bibitem[\protect\citeauthoryear{Hochreiter \& Schmidhuber}{Hochreiter \&
  Schmidhuber}{1997}]{hochreiter_long_1997}
Hochreiter S.,  Schmidhuber J.,  1997, \mn@doi [Neural Computation]
  {10.1162/neco.1997.9.8.1735}, 9, 1735

\bibitem[\protect\citeauthoryear{{Jaderberg}, {Simonyan}, {Zisserman}  \&
  {Kavukcuoglu}}{{Jaderberg} et~al.}{2015}]{2015arXiv150602025J}
{Jaderberg} M.,  {Simonyan} K.,  {Zisserman} A.,   {Kavukcuoglu} K.,  2015,
  arXiv e-prints, \href {https://arxiv.org/abs/1506.02025} {p.
  arXiv:1506.02025}

\bibitem[\protect\citeauthoryear{Jamal \& Bloom}{Jamal \&
  Bloom}{2020}]{jamal_neural_2020}
Jamal S.,  Bloom J.~S.,  2020, \mn@doi [The Astrophysical Journal Supplement
  Series] {10.3847/1538-4365/aba8ff}, 250, 30

\bibitem[\protect\citeauthoryear{Jayasinghe et~al.,}{Jayasinghe
  et~al.}{2018}]{jayasinghe_asas-sn_2018}
Jayasinghe T.,  et~al., 2018, \mn@doi [Monthly Notices of the Royal
  Astronomical Society] {10.1093/mnras/sty838}, 477, 3145

\bibitem[\protect\citeauthoryear{Jayasinghe et~al.,}{Jayasinghe
  et~al.}{2019}]{jayasinghe_asas-sn_2019}
Jayasinghe T.,  et~al., 2019, \mn@doi [Monthly Notices of the Royal
  Astronomical Society] {10.1093/mnras/stz844}, 486, 1907

\bibitem[\protect\citeauthoryear{Kim \& Bailer-Jones}{Kim \&
  Bailer-Jones}{2016}]{kim_package_2016}
Kim D.-W.,  Bailer-Jones C. A.~L.,  2016, \mn@doi [Astronomy \& Astrophysics]
  {10.1051/0004-6361/201527188}, 587, A18

\bibitem[\protect\citeauthoryear{Kim \& Brunner}{Kim \&
  Brunner}{2017}]{kim_stargalaxy_2017}
Kim E.~J.,  Brunner R.~J.,  2017, \mn@doi [Monthly Notices of the Royal
  Astronomical Society] {10.1093/mnras/stw2672}, 464, 4463

\bibitem[\protect\citeauthoryear{Kingma \& Ba}{Kingma \&
  Ba}{2014}]{kingma_adam:_2014}
Kingma D.~P.,  Ba J.,  2014, arXiv:1412.6980 [cs]

\bibitem[\protect\citeauthoryear{Krueger et~al.,}{Krueger
  et~al.}{2017}]{krueger_zoneout_2017}
Krueger D.,  et~al., 2017, arXiv:1606.01305 [cs]

\bibitem[\protect\citeauthoryear{{Kunder}, {Popowski}, {Cook}  \&
  {Chaboyer}}{{Kunder} et~al.}{2008}]{2008AJ....135..631K}
{Kunder} A.,  {Popowski} P.,  {Cook} K.~H.,   {Chaboyer} B.,  2008, \mn@doi
  [Astronomical Journal] {10.1088/0004-6256/135/2/631}, \href
  {https://ui.adsabs.harvard.edu/abs/2008AJ....135..631K} {135, 631}

\bibitem[\protect\citeauthoryear{Le, Jaitly  \& Hinton}{Le
  et~al.}{2015}]{le_simple_2015}
Le Q.~V.,  Jaitly N.,   Hinton G.~E.,  2015, arXiv:1504.00941 [cs]

\bibitem[\protect\citeauthoryear{LeCun, Bengio  \& Hinton}{LeCun
  et~al.}{2015}]{lecun_deep_2015}
LeCun Y.,  Bengio Y.,   Hinton G.,  2015, \mn@doi [Nature]
  {10.1038/nature14539}, 521, 436

\bibitem[\protect\citeauthoryear{Lecun, Bottou, Bengio  \& Haffner}{Lecun
  et~al.}{1998}]{lecun_gradient-based_1998}
Lecun Y.,  Bottou L.,  Bengio Y.,   Haffner P.,  1998, \mn@doi [Proceedings of
  the IEEE] {10.1109/5.726791}, 86, 2278

\bibitem[\protect\citeauthoryear{Lomb}{Lomb}{1976}]{lomb_least-squares_1976}
Lomb N.~R.,  1976, \mn@doi [Astrophysics and Space Science]
  {10.1007/BF00648343}, 39, 447

\bibitem[\protect\citeauthoryear{{Louppe}, {Cho}, {Becot}  \&
  {Cranmer}}{{Louppe} et~al.}{2019}]{2019JHEP...01..057L}
{Louppe} G.,  {Cho} K.,  {Becot} C.,   {Cranmer} K.,  2019, \mn@doi [Journal of
  High Energy Physics] {10.1007/JHEP01(2019)057}, \href
  {https://arxiv.org/abs/1702.00748} {2019, 57}

\bibitem[\protect\citeauthoryear{Mattheakis, Protopapas, Sondak, Di~Giovanni
  \& Kaxiras}{Mattheakis et~al.}{2020}]{mattheakis_physical_2020}
Mattheakis M.,  Protopapas P.,  Sondak D.,  Di~Giovanni M.,   Kaxiras E.,
  2020, arXiv:1904.08991 [physics]

\bibitem[\protect\citeauthoryear{McNamara, Clementini  \& Marconi}{McNamara
  et~al.}{2007}]{mcnamara_2007}
McNamara D.~H.,  Clementini G.,   Marconi M.,  2007, \mn@doi [The Astronomical
  Journal] {10.1086/513717}, 133, 2752

\bibitem[\protect\citeauthoryear{Minniti et~al.,}{Minniti
  et~al.}{2010}]{minniti_vista_2010}
Minniti D.,  et~al., 2010, \mn@doi [New Astronomy]
  {10.1016/j.newast.2009.12.002}, 15, 433

\bibitem[\protect\citeauthoryear{M\"oller \& de Boissi\`ere}{M\"oller \&
  de~Boissi\`ere}{2020}]{moller_supernnova_2020}
M\"oller A.,  de Boissi\`ere T.,  2020, \mn@doi [Monthly Notices of the Royal
  Astronomical Society] {10.1093/mnras/stz3312}, 491, 4277

\bibitem[\protect\citeauthoryear{Narayan et~al.,}{Narayan
  et~al.}{2018}]{narayan_machine-learning-based_2018}
Narayan G.,  et~al., 2018, \mn@doi [The Astrophysical Journal Supplement
  Series] {10.3847/1538-4365/aab781}, 236, 9

\bibitem[\protect\citeauthoryear{Naul, Bloom, P\'erez  \& van~der Walt}{Naul
  et~al.}{2018}]{naul_recurrent_2018}
Naul B.,  Bloom J.~S.,  P\'erez F.,   van~der Walt S.,  2018, \mn@doi [Nature
  Astronomy] {10.1038/s41550-017-0321-z}, 2, 151

\bibitem[\protect\citeauthoryear{North, Gauderon, Barblan  \& Royer}{North
  et~al.}{2012}]{north_vlt_2012}
North P.,  Gauderon R.,  Barblan F.,   Royer F.,  2012, \mn@doi [Astronomy and
  Astrophysics] {10.1051/0004-6361/200810284e}, 540, C1

\bibitem[\protect\citeauthoryear{Nun, Protopapas, Sim, Zhu, Dave, Castro  \&
  Pichara}{Nun et~al.}{2015}]{nun_fats_2015}
Nun I.,  Protopapas P.,  Sim B.,  Zhu M.,  Dave R.,  Castro N.,   Pichara K.,
  2015, arXiv:1506.00010 [astro-ph]

\bibitem[\protect\citeauthoryear{{Paczy\'nski}}{{Paczy\'nski}}{1997}]{1997eds..proc..273P}
{Paczy\'nski} B.,  1997, in {Livio} M.,  {Donahue} M.,   {Panagia} N.,  eds,
  The Extragalactic Distance Scale. p.~273

\bibitem[\protect\citeauthoryear{Richards et~al.,}{Richards
  et~al.}{2011}]{richards_machine-learned_2011}
Richards J.~W.,  et~al., 2011, \mn@doi [The Astrophysical Journal]
  {10.1088/0004-637X/733/1/10}, 733, 10

\bibitem[\protect\citeauthoryear{Scargle}{Scargle}{1982}]{scargle_studies_1982}
Scargle J.~D.,  1982, \mn@doi [The Astrophysical Journal] {10.1086/160554},
  263, 835

\bibitem[\protect\citeauthoryear{Shallue \& Vanderburg}{Shallue \&
  Vanderburg}{2018}]{shallue_identifying_2018}
Shallue C.~J.,  Vanderburg A.,  2018, \mn@doi [The Astronomical Journal]
  {10.3847/1538-3881/aa9e09}, 155, 94

\bibitem[\protect\citeauthoryear{Skowron et~al.,}{Skowron
  et~al.}{2019}]{Skowron478}
Skowron D.~M.,  et~al., 2019, \mn@doi [Science] {10.1126/science.aau3181}, 365,
  478

\bibitem[\protect\citeauthoryear{Tachibana, Graham, Kawai, Djorgovski, Drake,
  Mahabal  \& Stern}{Tachibana et~al.}{2020}]{tachibana_deep_2020}
Tachibana Y.,  Graham M.~J.,  Kawai N.,  Djorgovski S.~G.,  Drake A.~J.,
  Mahabal A.~A.,   Stern D.,  2020, arXiv:2003.01241 [astro-ph]

\bibitem[\protect\citeauthoryear{Torres \& Ribas}{Torres \&
  Ribas}{2002}]{torres_absolute_2002}
Torres G.,  Ribas I.,  2002, \mn@doi [The Astrophysical Journal]
  {10.1086/338587}, 567, 1140

\bibitem[\protect\citeauthoryear{Tsang \& Schultz}{Tsang \&
  Schultz}{2019}]{tsang_deep_2019}
Tsang B. T.-H.,  Schultz W.~C.,  2019, \mn@doi [The Astrophysical Journal]
  {10.3847/2041-8213/ab212c}, 877, L14

\bibitem[\protect\citeauthoryear{Udalski}{Udalski}{2003}]{udalski_optical_2003}
Udalski A.,  2003, Acta Astronomica, 53, 291

\bibitem[\protect\citeauthoryear{Wisdom, Powers, Hershey, Roux  \&
  Atlas}{Wisdom et~al.}{2016}]{wisdom_full-capacity_2016}
Wisdom S.,  Powers T.,  Hershey J.~R.,  Roux J.~L.,   Atlas L.,  2016,
  arXiv:1611.00035 [cs, stat]

\bibitem[\protect\citeauthoryear{Zhang \& Bloom}{Zhang \&
  Bloom}{2020}]{zhang_deepcr_2020}
Zhang K.,  Bloom J.~S.,  2020, \mn@doi [The Astrophysical Journal]
  {10.3847/1538-4357/ab3fa6}, 889, 24

\makeatother
\end{thebibliography}




\appendix

\section{PP-MNIST: Periodic Permuted MNIST}
\label{sec:PPMNIST}

\begin{figure*}
 \centering
 \includegraphics[width=\textwidth]{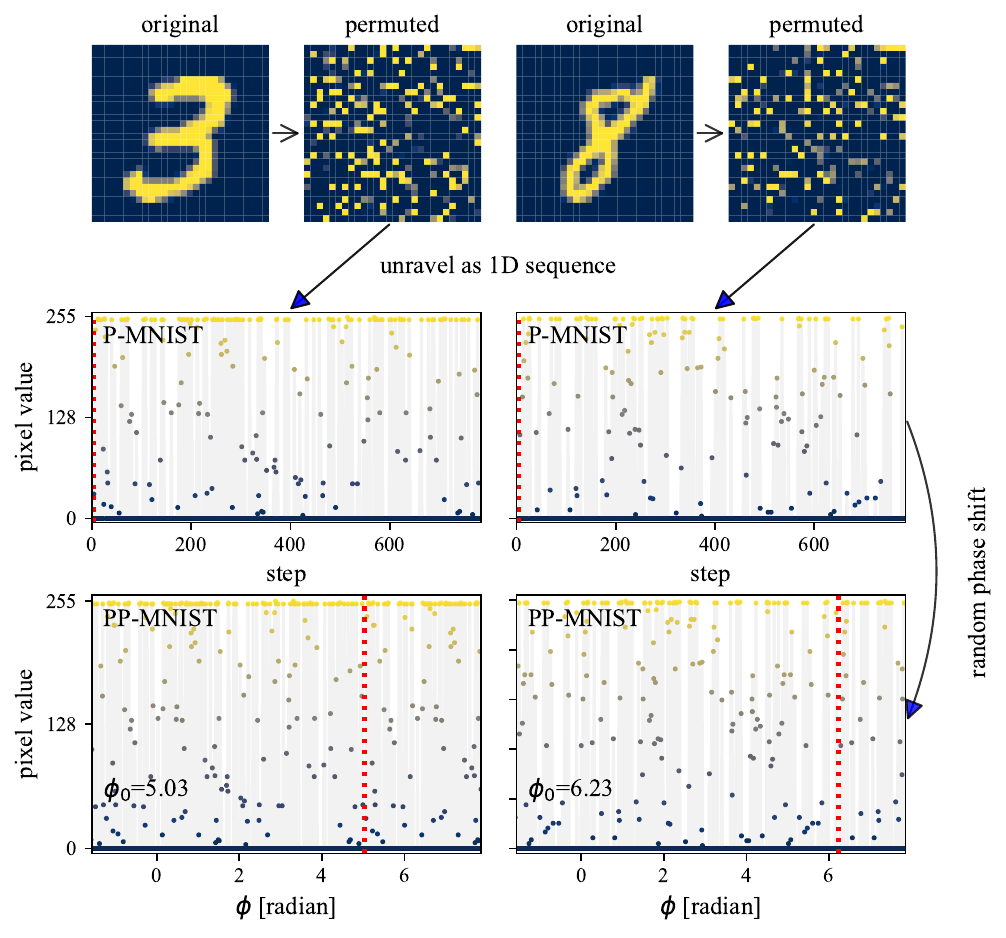}
 \caption{Construction of the PP-MNIST experiment. Pixel value is color coded with blue (0) transitioning to yellow (255). (top) The original 28$\times$28 MNIST image and the same image with a fixed pixel order permutation. (middle) The P-MNIST 1D sequence, where the red vertical dashed lines indicate that no phase-shift is applied. (bottom) The PP-MNIST 1D sequence, which is the same sequence as the middle row, but with a phase-shift. The vertical red dashed lines indicate the initial-phase of the PP-MNIST sequence, whose numerical value is indicated in the bottom left corners.}
 \label{fig:ppmnist}
\end{figure*}

To examine the effectiveness of the cyclic-permutation invariant networks for classification tasks in other domains, we have created an additional benchmarking dataset, ``periodic permuted MNIST'' (hereafter PP-MNIST), which is derived from the ``sequential MNIST'' and ``permuted MNIST'' classification tasks (Figure \ref{fig:ppmnist}). MNIST is a classic image dataset \citep{lecun_gradient-based_1998} consisting of 70,000 $28\times28$ images of hand-written digits in 10 classes (0 to 9). Under the sequential MNIST task, the 2D MNIST images are unwrapped into a 1D sequence of $L=784$. Sequential MNIST is frequently used to test a recurrent network's ability to retain long-range information \citep{le_simple_2015, wisdom_full-capacity_2016, krueger_zoneout_2017}. For the more challenging permuted MNIST (P-MNIST) task, a fixed random permutation is applied to each sequence \citep{le_simple_2015, wisdom_full-capacity_2016, krueger_zoneout_2017, arjovsky_unitary_2016} so that any spatial/temporal structure is removed. It has been shown in \cite{bai_empirical_2018} that TCNs outperform RNN baselines for both sequential MNIST and P-MNIST. Here, we introduce periodicity to P-MNIST by introducing a random cyclic-permutation to each P-MNIST sequence. Because any of the 784 locations could be the zero index after permutation, only the relative, cyclic ordering of the sequence remains meaningful. Just as the case of periodic variable star classification, doing so essentially wraps each P-MNIST sequence in a ring whereby the ``initial phase'' of the sequence becomes a nuisance parameter and is no longer relevant for the classification.

We test the iResNet and the iTCN against their non-invariant counterparts to show improvements enabled by cyclic-permutation invariance. A hyper-parameter search is done for iResNet/ResNet over depth (9, 10), initial hidden dimension (24, 48), maximum hidden dimension (120, 200), and for iTCN/TCN over depth (8, 9), kernel size (3, 7), and hidden dimension (24, 48, 96). We present the PP-MNIST test accuracies in Table \ref{table:ppmnist}. Both invariant networks outperform their non-invariant counterparts, especially in the case of the iTCN/TCN. The poor performance of TCN can be partially attributed to the exceptionally large (784) number of possible initial phases for each sequence, four times more than the $L=200$ sequences for periodic variable star classification. On the other hand, the regular ResNet performed relatively well. This is not surprising as the ResNet is by design different from the TCN --- the ResNet is based on localised feature extraction where features are condensed through \texttt{pooling} layers, but the TCN is a sequential model subject to the causal condition, which requires it to memorize features extracted in temporal order.

\begin{table}
 \caption{Periodic permuted MNIST (PP-MNIST) classification accuracies.}
 \begin{center}
   \begin{tabular}{*{4}{c}}
    \toprule
    iResNet& \textbf{96.0\%}&  iTCN& \textbf{94.8\%}\\
    ResNet& 95.1\%& TCN& 77.4\%\\
    \bottomrule
 
 \end{tabular}
 \end{center}
\label{table:ppmnist}
\end{table}

\section{Neural Network Hyper-Parameter Optimization}
\label{sec:hyperops}
We search for optimal hyper-parameters independently for each network and for each dataset with the validation set in a fixed train/validation/test split. For all networks, among possible combinations of input features: phase interval ($\Delta\pmb{\phi}$), magnitude ($\vect{m}$), magnitude change ($\Delta \vect{m}$), and gradient ($\Delta \vect{m}/\Delta \pmb{\phi}$), we find the combination of $(\Delta \pmb{\phi}, \vect{m})$ to yield the highest validation accuracy. For iTCN/TCN, we perform a hyper-parameter search over network depth (6, 7), hidden dimension (12, 24, 48), dropout (0, 0.15, 0.25), and kernel size (iTCN/TCN: 2, 3, 5). For iResNet/ResNet, we perform a grid search over initial hidden dimension (16, 32), maximum hidden dimension (32, 64), network depth (4, 5, 6), and kernel size (3, 5, 7). For GRU/LSTM, we search over network depth (2, 3), hidden dimension (12, 24, 48), and dropout rate (0, 0.15, 0.25). We find that a dropout rate of 0.15 works best for both GRUs and LSTMs across all three datasets, while no dropout works best for all other networks.

All networks are trained with the ADAM optimiser \citep{kingma_adam:_2014} with initial learning rates of 0.005, which are scheduled to decrease by a factor of 0.1 when training loss does not decrease by 10\% for 5 epochs. Models are saved at the best validation accuracy for testing.

\section{Data augmentation}
\label{sec:aug}
Both the semi-invariant iResNet and non-invariant baseline networks potentially benefit from data augmentation of the initial-phase during training. Using cyclic-permutations of the input sequence as training-time data augmentation, we trained iResNets and ResNets on the three datasets, after redoing hyperparameter optimisation. As seen in Table \ref{table:aug}, classification accuracies of both the iResNet and the ResNet are increased in most cases; the iResNets still hold a statistically significant advantage over the ResNets.

\begin{table}
 \caption{Classification accuracies for networks with and without data augmentation. Accuracies without data augmentation is identical to Table \ref{table:vs}.}
 \begin{center}
    \begin{tabular}[\textwidth]{*{4}{c}}
    \toprule
    Model & MACHO& OGLE-III& ASAS-SN\\
    \midrule
    \multicolumn{4}{c}{without phase data-augmentation}\\
    \midrule
    iResNet& $92.6\%\pm0.45\%$& $\textbf{93.7\%}\pm0.09\%$& $93.9\%\pm0.14\%$ \\
    ResNet& $92.1\%\pm0.34\%$& $93.4\%\pm0.11\%$& $93.2\%\pm0.22\%$ \\
    \textit{diff} & $({-0.48\%}_{-0.13\%}^{+0.05\%})$& $({-0.25\%}_{-0.02\%}^{+0.03\%})$& $({-0.64\%}_{-0.07\%}^{+0.01\%})$ \\
    \midrule
    \multicolumn{4}{c}{with phase data-augmentation}\\
    \midrule
    iResNet& $\textbf{92.9\%}\pm0.33\%$& $\textbf{93.7\%}\pm0.11\%$& $\textbf{94.4\%}\pm0.18\%$ \\
    ResNet& $92.4\%\pm0.29\%$& $93.5\%\pm0.12\%$& $94.2\%\pm0.23\%$ \\
    \textit{diff} & $({-0.39\%}_{-0.11\%}^{+0.08\%})$& $({-0.19\%}_{-0.04\%}^{+0.02\%})$& $({-0.32\%}_{-0.01\%}^{+0.14\%})$ \\
    \bottomrule
 \end{tabular}
 \end{center}
 \label{table:aug}
\end{table}


\bsp	
\label{lastpage}
\end{document}